\documentclass[aps,twocolumn,groupedaddress,amsmath,amssymb]{revtex4}
\usepackage{graphicx}
\usepackage{dcolumn}
\usepackage{bm}
\usepackage{amssymb}
\usepackage{amsmath}
\usepackage{epsfig}    
\usepackage{color}
\usepackage{slashed}
\usepackage{hhline}

\def\be{\begin{equation}}
\def\ee{\end{equation}}
\newcommand{\bea}{\begin{eqnarray}}
\newcommand{\eea}{\end{eqnarray}}
\newcommand{\nn}{\nonumber}

\begin{document}

 \begin{flushright} {KIAS-P18078, APCTP Pre2018 - 007}  \end{flushright}

\title{An inverse seesaw model with natural hierarchy at TeV scale }

\author{Takaaki Nomura}
\email{nomura@kias.re.kr}
\affiliation{School of Physics, KIAS, Seoul 02455, Republic of Korea}

\author{Hiroshi Okada}
\email{okada.hiroshi@apctp.org}
\affiliation{Asia Pacific Center for Theoretical Physics, Pohang, Gyeongbuk 790-784, Republic of Korea}

\date{\today}

\begin{abstract}
 We propose a new kind of inverse seesaw model without any additional symmetries.
Instead of the symmetries, we introduce several fermions and bosons with higher $SU(2)_L$ representations.
After formulating the Higgs sector and neutrino sector, we show that the cut-off energy, which is valid to our model, is at around TeV scale,
by examining behavior of $SU(2)_L$ gauge coupling.
Then we show the testability of our model at collider physics. 
 \end{abstract}
\maketitle

\section{Introductions}
One of the important topics in particle physics beyond the standard model (SM) is to explore mechanisms generating tiny neutrino masses
and its mixing.
Sometimes, one relies upon a mechanism in which the tininess originates from ultra high energy scale $\Lambda_H$ such as grand unified theory scale~($\sim10^{15}$ GeV) or string scale~($\sim10^{18}$ GeV), by embedding the SM gauge group into larger one. 
In this case, the neutrino mass may simply be realized by running a heavy field inside a diagram for the neutrino mass, therefore the neutrino mass is proportional to $\Lambda_{ew}/\Lambda_H$, where $\Lambda_{ew}$ is the electroweak scale$\sim$ 100GeV. These theories are elegant in a sense that the neutrino mass is directly reflected to the high energy scale, however the heavy particle cannot directly be tested by any current experiments such as the large hadron collider (LHC).
To achieve the detectability at the current experiments, a theory should be closed within TeV scale.
To realize such a low energy scenario, we apply higher multiplet fields under $SU(2)_L$ group instead of introducing any larger gauge groups
\footnote{There are several explanations to describe the tiny neutrino mass in a low energy scale. See, e.g., ref.~\cite{Ma:2006km}.}.
Once the neutrino mass is induced after a higher iso-spin scalar field developing  the vacuum expectation value (VEV), the order of neutrino mass is, at least, suppressed by two orders of magnitude compared to the origin of SM Higgs. This result follows from the bound on $\rho$ parameter that is describe by the mass ratio between the neutral gauge boson and charged gauge boson in the SM. The higher multiplet particles there exist, the more suppression the neutrino mass receives.
Thus, we realize the tiny neutrino mass with a natural hierarchy, by applying this feature.

To achieve this mechanism with more effective manner,  
an inverse seesaw model~\cite{Mohapatra:1986bd, Wyler:1982dd} is a promising candidate to explain the
miniscule neutrino masses with mild hierarchy of Majorana mass matrix for neutral Majorana fermions in a theory, and provides a lot of phenomenologies 
since the neutrino mass structure is more intricate than the other mechanisms such as canonical seesaw~\cite{Seesaw1, Seesaw2, Seesaw3, Seesaw4} and linear seesaw~\cite{Wyler:1982dd, Akhmedov:1995ip, Akhmedov:1995vm}.
In order to realize the inverse seesaw model, heavier neutral fermions with both chiralities have to be introduced.
In addition to these new fields, most of the cases, one also has to impose an additional symmetry such as (Non-)Abelian local(global) one
 to control the texture of neutral fermion mass matrix.

In this letter, we propose an inverse seesaw model without introducing any additional symmetries to the SM.
Instead, we introduce several fermions (quartet and septet) and bosons (quintet and quartet) with {\it higher $SU(2)_L$ representations}~\footnote{Several representative ideas along this line have been done in refs~\cite{Nomura:2018ibs, Nomura:2018cle, Kumericki:2012bh, Law:2013gma, Yu:2015pwa, Nomura:2016jnl,Nomura:2017abu, Wang:2016lve}}.
Due to such fields, behavior of $SU(2)_L$ gauge coupling $g_2$ blows up at TeV scale via renormalization group equation (RGE). 
It suggests that our model is tightly relevant at low energy scale, and the testability of our model is largely expected at various experiments such as the LHC, the ILC, and future colliders.

This letter is organized as follows.
In Sec. II, {we review our model and formulate the lepton sector. 
Then we discuss phenomenologies of neutrinos. In Sec. III we discuss an extra charged particles at collider experiments.
 Finally we devote the summary of our results and the conclusion.}

\section{Model setup and Constraints}
\begin{table}[t!]
\begin{tabular}{|c||c|c|c|c||c|c|c|}\hline\hline  
& ~$L_L^a$~& ~$e_R^a$~& ~$\psi^a$~& ~$\Sigma_R^a$~& ~$H_2$~& ~$H_4$~& ~$H_5$~\\\hline\hline 
$SU(2)_L$   & $\bm{2}$  & $\bm{1}$  & $\bm{4}$  & $\bm{7}$ & $\bm{2}$   & $\bm{4}$ & $\bm{5}$   \\\hline 
$U(1)_Y$    & $-\frac12$  & $-1$ & {-$\frac32$}  & $0$  & $\frac12$ & {$\frac32$}  &{$2$} \\\hline
\end{tabular}
\caption{Charge assignments of the our lepton and scalar fields
under $SU(2)_L\times U(1)_Y$, where the upper index $a$ is the number of family that runs over 1-3 and
all of them are singlet under $SU(3)_C$. }\label{tab:1}
\end{table}

In this section we formulate our model.
For the fermion sector, we introduce three families of vector-like fermions $\psi$ with $(\bm{4},-3/2)$ charge under the $SU(2)_L\times U(1)_Y$ gauge symmetry, and right-handed fermions $\Sigma_R$ with $(\bm{7},0)$ charge under the same gauge symmetry.
%
For the scalar sector, we add quartet and quintet scalar fields $H_4$ and $H_5$ with respectively {$3/2$} and $2$ charges under the $U(1)_Y$ gauge symmetry, where SM-like Higgs field is identified as $H_2$.
Here we denote each vacuum expectation value{(VEV)} of scalar fields to be $\langle H_i\rangle\equiv v_{i}/\sqrt2$ ($i=2,4,5$)
that is arisen after the electroweak spontaneously symmetry breaking.
All the field contents and their assignments are summarized in Table~\ref{tab:1}, where the quark sector is exactly same as the one of the SM and omitted.
The renormalizable Yukawa Lagrangian under these symmetries is given by
\begin{align}
-{\cal L_\ell}
& =  y_{\ell_{aa}} \bar L^a_L H_2 e^a_R  +  y_{D_{ab}} [ \bar L^a_L H_5^* (\psi^c_L)^b ]  \nn\\
& +  f_{{L}_{ab}} [\bar \psi^{a}_L  H_4^* \Sigma^b_R]
+ f_{{R}_{ab}} [(\bar\psi^c_R)^a  H_4\Sigma^b_R] \nn\\
&+M_{aa} \bar \psi^a_R \psi_L^a+ M_{\Sigma_{aa}}(\bar \Sigma^c_R)^a \Sigma_R^{a}
+ {\rm h.c.}, \label{Eq:yuk}
\end{align}
where $SU(2)_L$ index is omitted assuming it is contracted to be gauge invariant, 
and upper indices $(a,b)=1$-$3$ are the number of families, and $y_\ell$, $M$, and $M_\Sigma$ are assumed to be  diagonal matrix with real parameters.
After spontaneous symmetry breaking we obtain mass matrices $m_\ell=y_\ell v/\sqrt2$ and $m_D=y_D v_5/\sqrt2$.   

{
\noindent \underline{\it Scalar potential and VEVs}:
The scalar potential in our model is given by
\begin{align}
{\cal V} = & -\mu_h^2 |H_2|^2 + M_4^2 |H_4|^2 + M_5^2 |H_5|^2  + \lambda_H |H_2|^4 \nonumber \\
& + \mu [H_4 H_2  H_5^* +h.c] + \lambda_0 [H_4^* H_2 H_2 H_2 + h.c.] \nonumber \\
& + {\cal V}_{\rm trivial},
\end{align}
where ${\cal V}_{\rm trivial}$ indicates other trivial 4-point terms and $SU(2)_L$ indices are implicitly contracted in the second line to be gauge invariant.
Applying condition $\partial {\cal V}/\partial v_i = 0$, we obtain the VEVs as 
\begin{align}
v_2 \sim \sqrt{\frac{\mu^2_h}{\lambda_H}}, \quad v_4 \sim \frac{\lambda_0 v^3 }{M_4^2}, \quad v_5 \sim \frac{\mu v_4 v}{M_5^2},
\end{align}
where we have used $v_4, v_5 \ll v_2$.
Thus $v_4$ and $v_5$ can be naturally $\mathcal{O}(1)$ GeV scale if $M_4$ and $M_5$ are TeV scale.
}

\noindent \underline{\it $\rho$ parameter}:
The VEVs of $H_4$ and $H_5$ are restricted by the $\rho$-parameter at tree level  that is given by 
\begin{align}
\rho\approx \frac{v_2^2+7 v_4^2+10 v_5^2}{v_2^2+  v_4^2 + 2 v_5^2},
\end{align}
where the experimental value is given by $\rho=1.0004^{+0.0003}_{-0.0004}$ at $2\sigma$ confidence level~\cite{pdg}.
On the other hand, we have $v_{SM}=\sqrt{v_2^2+7 v_4^2+10 v_5^2}\simeq v_2\approx$246 GeV. Therefore $v_4$ and $v_5$ are restricted via the constraint of $\rho$ parameter.
Here, we take these VEVs to be $v_2\approx$245.9 GeV, $v_4\approx$1.67 GeV, and $v_5\approx$1.72 GeV, which are typical scale for the VEVs satisfying the constraint.
%

{
\noindent \underline{\it Exotic particles} :
The scalars and fermions with large $SU(2)_L$ multiplet provide exotic charged particles.
Here we write components of multiplets as
\begin{align}
& H_4 = (\phi_4^{+++}, \phi_4^{++}, \phi_4^{+}, \phi_4^0)^T, \\
& H_5 = (\phi_5^{++++}, \phi_5^{+++}, \phi_5^{++}, \phi_5^{+}, \phi_5^0)^T, \\
& \psi_{L(R)} = (\psi^{0}, \psi^{-}, \psi^{--}, \psi^{---})^T_{L(R)}, \\
& \Sigma_R = (\Sigma^{+++}, \Sigma^{++}, \Sigma^{+}, \Sigma^0, \Sigma^-, \Sigma^{--}, \Sigma^{---})_R^T. 
\end{align}
The masses of components in $H_4$ and $H_5$ are respectively given by $\sim M_4$ and $\sim M_5$ since $v_{4,5} \ll M_{4,5}$.
The charged components in $\psi_{L(R)}$ have Dirac mass $M$ and neutral component is discussed with neutrino sector below.
The septet fermion mass is $M_\Sigma$ and charged components have Dirac mass term constructed by pairs of positive-negative charged components in the multiplet.
Charged particles in the same multiplet have degenerate mass at tree level which will be shifted at loop level~\cite{Cirelli:2005uq}. }

\begin{figure}[tb]
\begin{center}
\includegraphics[width=5.0cm]{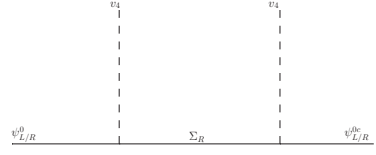}
\caption{Feynman diagram to generate the masses of $\mu_{L/R}$.}
\label{fig:mu_mass}
\end{center}\end{figure}

\noindent \underline{\it Neutrino sector}:
After the spontaneously symmetry breaking, neutral fermion mass matrix in basis of $(\nu_L,\psi_R^{0c},\psi_L^0)^T$ is given by
\begin{align}
M_N
&=
\left[\begin{array}{ccc}
0 & 0 & m_D^*  \\ 
0 & \mu_R^* & M \\ 
m_D^\dag  & M^T & \mu_L^* \\ 
\end{array}\right],
\end{align}
where $\mu_{L/R}$ is given by $v_4^2 f_{L/R}M^{-1}_{\Sigma}f^T_{L/R}$ on the analogical manner of seesaw mechanism,
as shown in Fig.~\ref{fig:mu_mass}.
Then the active neutrino mass matrix can approximately be found as
\begin{align}
m_\nu\approx m_D^* M_{}^{-1} \mu_R^* (M_{}^T)^{-1} m_D^\dag,
\end{align}
where $\mu_{L/R}< m_D\ll M_{}$ is naturally expected due to the constraint of $\rho$ parameter and seesaw-like mechanism of $\mu_{R/L}$
\footnote{These hierarchies could be explained by several mechanisms such as radiative models~\cite{Dev:2012sg, Dev:2012bd, Das:2017ski} and effective models with higher order terms \cite{Okada:2012np}.}.
{We thus obtain correlation among size of neutrino mass and other mass parameters such that
\begin{equation}
m_\nu \sim \frac{v_4}{M_\Sigma} \left( \frac{v_5}{M} \right)^2 f_R^2 y_D^2 v_4.
\end{equation}
Note that $M_\Sigma$ and $M$ cannot be much larger than TeV scale, since $v_4$ and $v_5$ are GeV scale requiring perturbative limit for Yukawa coupling constants.}
The neutrino mass matrix is diagonalized by unitary matrix $U_{MNS}$; $D_\nu= U_{MNS}^T m_\nu U_{MNS}$, where $D_\nu\equiv {\rm diag}(m_1,m_2,m_3)$.
{Here we apply a convenient method to reproduce neutrino oscillation data as follows~\cite{Nomura:2018mwr}:}
\begin{align}
m_D^* \approx U_{MNS}^* \sqrt{D_\nu} O_{mix} \sqrt{I_N} (L^T_N)^{-1}.
\end{align}
Here $O_{mix}$ is an arbitrary 3 by 3 orthogonal matrix with complex values, $I_N$ is a diagonal matrix, and $L_N$ is a lower unit triangular~\cite{Baek:2017qos}, which can uniquely be decomposed to be $M_{}^{-1} \mu_R^* (M_{}^T)^{-1}=L_N^T I_N L_N$,
since it is symmetric matrix. Note here that all the components of $m_D$ should not exceed {${\cal O}$(1)} GeV, once perturbative limit of $y_D$ is taken to be 1.

\noindent \underline{\it Non-unitarity}:
{Constraint of non-unitarity should always be taken into account in case of larger neutral mass matrix whose components are greater than three by three, since experimental neutrino oscillation results suggest nearly unitary.
In case of the inverse seesaw, when non-unitarity matrix $U'_{MNS}$ is defined, one can typically parametrize it by the following form:}
\begin{align}
U'_{MNS}\equiv \left(1-\frac12 FF^\dag\right) U_{MNS},
\end{align}
where $F\equiv  M_{}^{-1} m_D^*$ is a hermitian matrix, and $U'_{MNS}$ represents the deviation from the unitarity. 
%
{Considering several experimental bounds~\cite{Fernandez-Martinez:2016lgt},
one finds the following constraints~\cite{Agostinho:2017wfs}:}
\begin{align}
|FF^\dag|\le  
\left[\begin{array}{ccc} 
2.5\times 10^{-3} & 2.4\times 10^{-5}  & 2.7\times 10^{-3}  \\
2.4\times 10^{-5}  & 4.0\times 10^{-4}  & 1.2\times 10^{-3}  \\
2.7\times 10^{-3}  & 1.2\times 10^{-3}  & 5.6\times 10^{-3} \\
 \end{array}\right].
\end{align} 
Once we conservatively take $F\approx 10^{-5}$, we find {$\mu_{R}\approx$1-10 GeV to satisfy the typical neutrino mass scale, which could be easy task.}

\begin{figure}[tb]
\begin{center}
\includegraphics[width=7.0cm]{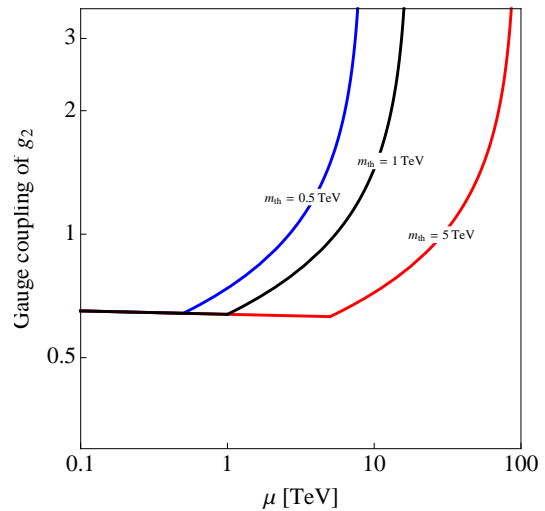}
\caption{The running of $g_2$ in terms of a reference energy of $\mu$.}
\label{fig:rge}
\end{center}\end{figure}

\noindent \underline{\it Beta function of $SU(2)_L$ gauge coupling $g_2$}
\label{beta-func}
Here it is worth  discussing the running of gauge coupling of $g_2$ in the presence of several new multiplet fields of $SU(2)_L$~\footnote{The gauge coupling of $U(1)_Y$ is relevant up to Planck scale.}.
The new contribution to $g_2$ for an $SU(2)_L$ quartet fermion(boson) $\psi(H_4)$, septet fermion $\Sigma_R$, and quintet boson $H_5$ are respectively given by
\begin{align}
 \Delta b^{\psi}_{g_2}=\frac{10}{3}, \ \Delta b^{\Sigma_R}_{g_2}=\frac{56}{3},\
  \Delta b^{H_4}_{g_2}=\frac{5}{3}, \ \Delta b^{H_5}_{g_2}=\frac{10}{3}  .
\end{align}
Then one finds the energy evolution of the gauge coupling $g_2$ as~\cite{Nomura:2017abu, Kanemura:2015bli}
\begin{align}
&\frac{1}{g^2_{g_2}(\mu)}=\frac1{g^2(m_{in})}-\frac{b^{SM}_{g_2}}{(4\pi)^2}\ln\left[\frac{\mu^2}{m_{in}^2}\right]
-\theta(\mu-m_{th}) \nn\\
&\times \frac{N_f( \Delta b^{\psi}_{g_2}+ \Delta b^{\psi}_{g_2})+\Delta b^{H_4}_{g_2}+\Delta b^{H_5}_{g_2}}{(4\pi)^2}\ln\left[\frac{\mu^2}{m_{th}^2}\right],
\label{eq:rge_g}
\end{align}
where $N_f=3$ is the number of $\psi$ and $\Sigma_R$, $\mu$ is a reference energy, $b^{SM}_{g_2}=-19/6$, and we assume $m_{in}(=m_Z) < m_{th}$ with $m_{th}$ being threshold masses of exotic fermions and bosons.
The resulting flow of ${g_2}(\mu)$ is then given by the Fig.~\ref{fig:rge}.
This figure shows that $g_2$ is relevant up to the mass scale $\mu={\cal O}(10)$ TeV in case of $m_{th}=$500 GeV,
while $g_2$ is relevant up to the mass scale $\mu={\cal O}(100)$ TeV in case of $m_{th}=$5000 GeV.
Thus our theory does not spoil, as far as we work on at around the scale of TeV.

\section{Collider physics}

\begin{figure}[tb]\begin{center}
\includegraphics[width=7.0cm]{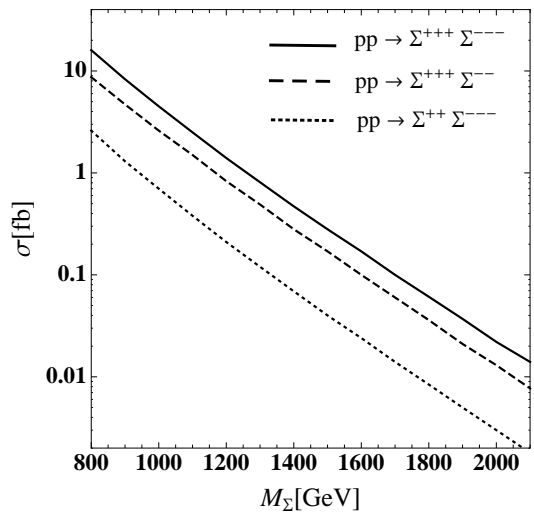}
\caption{The cross sections for processes $p p \to Z/\gamma \to \Sigma^{+++} \Sigma^{---}$ and $p p \to W^\pm \to \Sigma^{+++(++)} \Sigma^{--(---)}$ as a function of septet fermion mass.
}   \label{fig:CXsigma}\end{center}\end{figure}

Here let us discuss collider physics of our model.
We have rich phenomenology at collider experiments since there are many exotic charged particles from large $SU(2)$ multiplet scalars and fermions.
As the most specific signature we focus on the production of triply charged lepton $\Sigma^{\pm \pm \pm}$ and its decay at the LHC.
The gauge interactions associated with triply charged lepton are obtained as
\begin{align}
\bar \Sigma_R i \gamma^\mu D_\mu \Sigma_R \supset & \bar \Sigma^{+++} \gamma^\mu \left( 3 g_2 c_W  Z_\mu + 3 e A_\mu \right) \Sigma^{+++} \nonumber \\ 
 &+ \sqrt{3} g_2 \bar \Sigma^{+++} \gamma^\mu W^+_\mu \Sigma^{++} + h.c. ,
\end{align}
where $c_W = \cos \theta_W$ with Weinberg angle $\theta_W$ and $e$ is the electromagnetic coupling: covariant derivative for septet can be referred to ref.~\cite{Nomura:2016jnl}.
Then we estimate cross sections for triply charged lepton production processes using 
{\it CalcHEP}~\cite{Belyaev:2012qa} by use of the CTEQ6 parton distribution functions (PDFs)~\cite{Nadolsky:2008zw}, implementing relevant interactions.
In Fig.~\ref{fig:CXsigma}, we show production cross section for triply charged lepton as a function of its mass; pair production $pp \to \Sigma^{+++} \Sigma^{---}$ 
and associate productions $p p \to \Sigma^{+++(++)} \Sigma^{--(---)}$ at the LHC 13 TeV.
The cross section for pair production is the largest one and larger than $1$ fb for 1 TeV mass due to large charge.

The triply charged lepton can decay via Yukawa coupling in Eq.~(\ref{Eq:yuk}) as $\Sigma^{+++} \to \phi_4^{Q_\phi} \psi^{Q_\psi}$ where $Q_{\phi(\psi)}$ is the electric charge of $\phi(\psi)$ with $Q_\phi + Q_\psi =3$, 
and we assume exotic scalars are lighter than exotic fermions in our discussion.
In addition, $\psi^{Q_\psi}$ decays as $\psi_{Q_\psi} \to \phi_5^{Q'_\phi} \ell^+(\nu)$ with $Q_\psi = Q'_\phi +1(0)$.
There are several decay modes for exotic charged leptons due to combination of charges in final states which have similar size of branching ratio (BR).
Here we discuss the representative decay chain: 
\begin{align}
\Sigma^{+++} \to \phi_4^{++} \psi^+ \to  \phi_4^{++} \phi_5^0 \ell^+ \to W^+ W^+ Z Z \ell^+,
\end{align}
where $\phi_4^{++}$ and $\phi_5^0$ decay into $W^+W^+$ and $ZZ$ via gauge interaction~\cite{Nomura:2017abu}: 
\begin{align}
& (D_\mu H_4)^\dagger (D^\mu H_4) \supset \sqrt{\frac{3}{2}} v_4 g_2^2 W^\pm_\mu W^{\pm \mu} \phi_4^{\mp \mp} \\
& (D_\mu H_5)^\dagger (D^\mu H_5) \supset \frac{1}{8} \frac{g_2^2}{c_W^2} v_5 \phi_5^0 Z_\mu Z^\mu.
\end{align}
Note that BRs for $\phi_4^{\pm \pm} \to W^\pm W^\pm$ and $\phi_5^0 \to ZZ$ are dominant when $v_4 \sim v_5 \sim 1$ GeV.
When $W^+$ decays into leptons and $Z$ decays into jets we obtain signal of three same sign charged leptons with jets and missing transverse energy, 
which provides products of BRs; $BR(W^+ \to \ell^+ \nu)^2 BR(Z \to q \bar q)^2 \sim 0.02$ with $\ell = \mu, e$.
Thus we can obtain $\sim 6 (60)$ signal events containing three same sign charged leptons for integrated luminosity of 300(3000) fb$^{-1}$ when products of $\Sigma^{\pm \pm \pm}$ production cross section and $BR(\phi_4^{\pm \pm} \psi^\pm) BR(\psi^\pm \to \phi_5^0 \ell^\pm)$ is around 1~fb. This size of cross section can be obtained for $M_\Sigma \sim 1$ TeV where we show the expected number of event in Table.~\ref{tab:CX} for several values of $M_\Sigma$ considering two or one pair of same sign $W$ boson decaying into leptons.
We find that number of events tends to be too small when both same sign $W$ boson pairs $W^+W^+(W^-W^-)$ decay into leptons although signal will be very clear.
Thus signal of three same sign charged lepton with jets and $\slashed{E}_T$ can be good target in searching for the signature of our model.
We expect sizable discovery significance even if number of signal events is less than $10$ since the SM background is very small for three same sign charged lepton signals.

\begin{table}[t!]
\begin{tabular}{|c||cc|}\hline\hline  
Signal & $3 \ell^+ 3 \ell^- 8 j \slashed{E}_T$~ & ~$3 \ell^\pm  12j \slashed{E}_T$~ \\\hline\hline 
$\sigma \cdot BR (M_\Sigma = 0.8 \ {\rm TeV})$ [fb] & 0.73 (7.3) & 7.2 (72)  \\ \hline 
$\sigma \cdot BR (M_\Sigma = 1.0 \ {\rm TeV})$ [fb] & 0.21 (2.1) & 2.0 (20)  \\ \hline 
$\sigma \cdot BR (M_\Sigma = 1.2 \ {\rm TeV})$ [fb] & 0.064 (0.64) & 0.63 (6.3)  \\ \hline 
\end{tabular}
\caption{Number of expected signal events for two final state from $pp \to \Sigma^{+++} \Sigma^{---}$ production where integrated luminosity is taken as 300(3000) fb$^{-1}$. }
\label{tab:CX}
\end{table}

\section{Summary and Conclusions}
We have constructed an inverse seesaw model with large $SU(2)_L$ multiplet fields in which we have formulated the neutrino mass matrix to reproduce current neutrino oscillation data, satisfying $\rho$ parameter and non-unitarity bound.
We have also checked the relevant energy scale of our theory via RGE of $SU(2)_L$ gauge coupling $g_2$ that gives the most stringent constraint. 
Then we have analyzed collider physics focusing on triply charged lepton production at the LHC as a representative process of our model and show a possibility of detection.
We have found specific signal of the triply charged lepton as three same sign charged leptons with jets and missing transverse momentum.
The number of events of the signal can be detectable level with integrated luminosity 300(3000) fb$^{-1}$ when triply charged lepton mass is around 1 TeV.
More detailed analysis will be given elsewhere.
\section*{Acknowledgments}
This research is supported by the Ministry of Science, ICT \& Future Planning of Korea, the Pohang City Government, and the Gyeongsangbuk-do Provincial Government (H. O.).
H. O. is sincerely grateful for KIAS and all the members.

\end{document}